\documentclass{interspeech2025}
\usepackage{amsmath,graphicx,xcolor,lipsum}
\usepackage{pifont}
\newcommand\blfootnote[1]{%
  \begingroup
  \renewcommand\thefootnote{}\footnote{#1}%
  \addtocounter{footnote}{-3}%
  \endgroup
}
\def\L{{\cal L}}
\interspeechcameraready

\title{Feature Importance across Domains for Improving Non-Intrusive Speech Intelligibility Prediction in Hearing Aids}

\author[affiliation={1}]{Ryandhimas E.}{Zezario}
\author[affiliation={2}]{Sabato M.}{Siniscalchi}
\author[affiliation={3}]{Fei}{Chen}
\author[affiliation={1}]{Hsin-Min}{Wang}
\author[affiliation={1}]{Yu}{Tsao}

\affiliation{}{Academia Sinica}{Taiwan}
\affiliation{}{University of Palermo}{Italy}
\affiliation{}{Southern University of Science and Technology}{China}
\email{\{ryandhimas, yu.tsao\}@citi.sinica.edu.tw}

\begin{document}

\maketitle

\begin{abstract}
Given the critical role of non-intrusive speech intelligibility assessment in hearing aids (HA), this paper enhances its performance by introducing Feature Importance across Domains (FiDo). We estimate feature importance on spectral and time-domain acoustic features as well as latent representations of Whisper. Importance weights are calculated per frame, and based on these weights, features are projected into new spaces, allowing the model to focus on important areas early. Next, feature concatenation is performed to combine the features before the assessment module processes them. Experimental results show that when FiDo is incorporated into the improved multi-branched speech intelligibility model MBI-Net+, RMSE can be reduced by 7.62\% (from 26.10 to 24.11). MBI-Net+ with FiDo also achieves a relative RMSE reduction of 3.98\% compared to the best system in the 2023 Clarity Prediction Challenge. These results validate FiDo’s effectiveness in enhancing neural speech assessment in HA.

%\noindent\textbf{Index Terms}: speech intelligibility, hearing aid, hearing loss, weak supervision, feature importance

\end{abstract}

\section{Introduction}
\blfootnote{The work of Fei Chen was supported by the National Natural Science Foundation of China (Grant No. 62371217).}
Assessing speech intelligibility is an important factor in evaluating the performance of various speech processing applications, such as speech enhancement \cite{loizou2007speech, 10248166} and hearing aid (HA) devices \cite{barker22_interspeech, 10362991}. The most direct and reliable assessment method is for human listeners to assess how many words they can correctly recognize from a given audio sample. However, despite the reliability of human-based assessment scores, cost and practicality remain significant challenges in obtaining unbiased assessment scores. Therefore, more practical assessment methods that mimic human-based assessments are needed.

Previously, most traditional methods relied on signal processing and psychoacoustic principles to replicate human auditory perception for intelligibility assessment. Well-known methods include the speech intelligibility index (SII) \cite{ref_36}, extended SII (ESII) \cite{ref_37}, speech transmission index (STI) \cite{ref_38}, short-time objective intelligibility (STOI) \cite{ref_39}, modified binaural short-time objective intelligibility (MBSTOI) \cite{ANDERSEN20181}, and hearing aid speech perception index (HASPI) \cite{katehaspi}. Despite the notable assessment performance of these methods, the availability of clean references is crucial to obtain more accurate assessment scores, which limits their practicality in situations where clean audio is not available as a ground-truth reference.

Recently, with the emergence of deep learning models and the availability of large-scale human assessment labels, researchers have begun to use deep learning for speech assessment \cite{ssl-mos, mosa-net, zezario2024studyincorporatingwhisperrobust, yang22o_interspeech, cuervo2024speech, mogridge2024nonintrusive}, with the goal of replicating human assessment behavior and improving the practicality of assessment methods. Interestingly, deep learning models have shown great potential in assessing speech quality and intelligibility in a non-intrusive manner, and most current deep learning-based speech assessment systems integrate large pre-trained models (such as self-supervised learning (SSL) models \cite{ssl-mos, mosa-net, yang22o_interspeech} or Whisper \cite{zezario2024studyincorporatingwhisperrobust, cuervo2024speech, mogridge2024nonintrusive}) and utilize ensemble learning \cite{yang22o_interspeech, cuervo2024speech, mogridge2024nonintrusive}. 

With two Clarity Prediction Challenges \cite{10094918, barker2022is}, there is growing interest in deploying non-intrusive speech intelligibility prediction models for hearing aids (HA) \cite{cuervo2024speech, mogridge2024nonintrusive, zezario24_interspeech, chiang2023multiobjective, tu22_interspeech, MAWALIM2023109663}. In the first Clarity Prediction Challenge \cite{barker2022is}, the best-performing non-intrusive system combines hidden layer representations of automatic speech recognition (ASR) \cite{tu22_interspeech} and hidden layers of SSL models \cite{zezario2022mbi} to achieve robust performance. In the more recent challenge \cite{10094918}, the use of Whisper-generated acoustic features proved to be more effective in improving robustness, as the three best non-intrusive models all employed Whisper to generate acoustic features \cite{cuervo2024speech, mogridge2024nonintrusive, zezario24_interspeech}. Moreover, we note that the robustness of acoustic features plays a crucial role in achieving higher prediction performance. Therefore, further research on the most effective strategies for extracting acoustic features is necessary.

Inspired by the concept of estimating feature importance in computer vision \cite{9709938}, which has been shown to be effective in helping models focus on important regions from the early stages of the training process, we aim to adopt this concept and propose feature importance across domains (FiDo). To demonstrate proof-of-concept of our approach, we aim to incorporate the FiDo approach into MBI-Net+ \cite{zezario24_interspeech}. We choose MBI-Net+ as our base model because of its simple module implementation and its excellent performance in the latest Clarity Prediction Challenge. In our implementation, FiDo adopts a different feature extraction strategy compared to the original cross-domain feature extraction of MBI-Net+. First, we introduce feature-based concatenation when processing cross-domain acoustic features. This ensures that we maintain the original number of frames, thus avoiding an excessive number of frames when concatenating in the temporal dimension, as implemented in the original MBI-Net+. Second, we introduce multi-head self-attention (MHSA) \cite{NIPS2017_3f5ee243} with cross-domain feature concatenation to effectively capture the importance of acoustic features. Since FiDo utilizes three different types of acoustic features: spectral (PS), filterbank (FB), and Whisper (WS) features, we first concatenate the MHSA outputs of PS and FB along the feature dimension. These features are then processed through a convolutional neural network and combined with the Whisper features processed using MHSA and an adapter layer. The final output of the FiDo feature extraction method is the combination of these two feature outputs. For the assessment module, we follow the original MBI-Net+ architecture, which includes BLSTM with attention, task-specific modules for estimating intelligibility, HASPI, and HA classes. Experimental results show that integrating FiDo into MBI-Net+ results in an relative RMSE reduction of 7.62\% (from 26.10\% to 24.11\%). Furthermore, MBI-Net+ with FiDo achieves an relative RMSE reduction of 3.98\% over the best-performing system  in the 2023 Clarity Prediction Challenge. These results confirm FiDo’s effectiveness in improving neural speech intelligibility assessment in HA.

The remainder of this paper is organized as follows. Section II presents the proposed FiDo method. Section III describes the experimental setup and results. Finally, Section IV provides conclusions and future work.
%\graphicspath{ {./images/} }
%\begin{figure}[t]
%\centering
%\includegraphics[width=8cm]{Figure/MBI-Net_.pdf} 
%\caption{Architecture of the MBI-Net+ model.} 
%\label{fig:MOSA-Net}
%\end{figure}

\graphicspath{ {./images/} }
\begin{figure}[t]
\centering
\includegraphics[width=8.25cm]{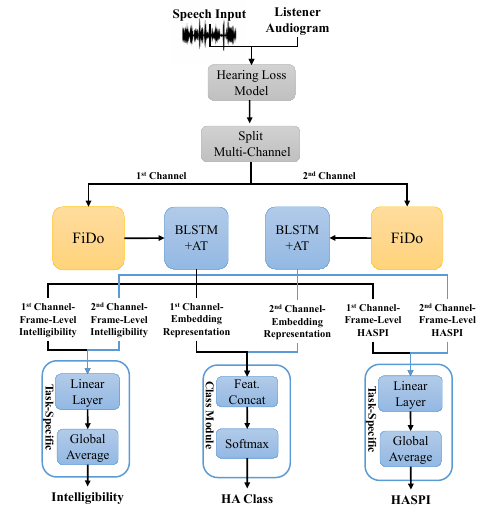} 
\caption{Architecture of the MBI-Net+ model with FiDo.} 
\label{fig:MBI_Net_FiDO}
\end{figure}

%\graphicspath{ {./images/} }
%\begin{figure}[t]
%\centering
%\includegraphics[width=5.5cm]{Figure/FIDO_detailed.pdf} 
%\caption{Illustration of extracting cross-domain features and estimating frame-level intelligibility scores using the CNN-BLSTM+AT architecture.} 
%\label{fig:FiDO}
%\end{figure}

\section{FiDo}

The motivation behind FiDo is to estimate attention weights immediately after feature extraction, with the expectation that it can improve prediction accuracy by making the model focus on salient acoustic features from the beginning. Unlike existing speech assessment methods that apply attention mechanisms at deeper layers \cite{mogridge2024nonintrusive, zezario24_interspeech, cuervo2024speech}, FiDo introduces early attention modulation to refine feature representations before higher-level processing. The overall integration of FiDo and MBI-Net+ is shown in Fig. \ref{fig:MBI_Net_FiDO}.

\begin{figure}[t]
\centering
\includegraphics[width=8.2cm]{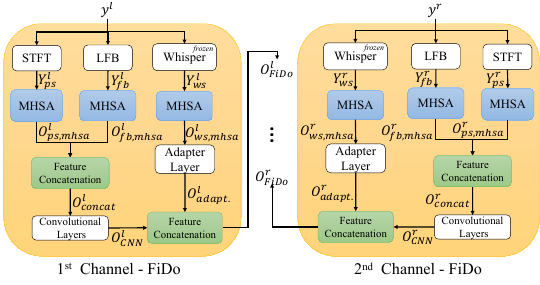} 
\caption{A detailed procedure of FiDo for both branches.} 
\label{fig:detailfido}
\end{figure}

The FiDo acoustic feature extraction approach is applied immediately after splitting the simulated audio (compensated for specific audiogram characteristics) into two separate channels ($y^l, y^r$). As illustrated in Fig. \ref{fig:detailfido}, two distinct FiDo modules are used to process each audio channel input. In the detailed process, given the simulated audio input of the left channel $y^l$ (the steps for the right channel are the same), the corresponding features (PS, FB, and WS) are extracted as follows:  

\graphicspath{ {./images/} }

\begin{equation}
\begin{array}{c}
Y^l_{ps} = \text{STFT}(y^l), \\
Y^l_{fb} = \text{LFB}(y^l), \\
Y^l_{ws} = \text{Whisper}(y^l). \\
\end{array}
\label{eq:extract}
\end{equation}
To ensure that the three types of features extracted from an audio input have the same number of time frames, we perform zero padding to 7 seconds on each input waveform and adjust Whisper's chunk size to 7-seconds. In addition, for \( Y^l_{ps} \) and \( Y^l_{fb} \), we apply the same configuration to the STFT operation and the SincNet \cite{sincnet} layer of learnable filterbank (LFB) to be consistent with Whisper's frame rate.

After obtaining the three corresponding features—$Y^l_{ps} \in \mathbb{R}^{T \times (d_{ps})}$, $Y^l_{fb} \in \mathbb{R}^{T \times (d_{fb})}$, and $Y^l_{ws} \in \mathbb{R}^{T \times (d_{ws})}$—where $T$ represents the number of time frames, and $d_{ps}$, $d_{fb}$, and $d_{ws}$ denote the feature dimensions, the process of estimating feature importance is formulated as follows:

\begin{equation}
\begin{array}{c}
Q^l_{ps} = Y^l_{ps} W_{ps,Q}^{l}, \quad K^l_{ps} = Y^l_{ps} W_{ps,K}^{l}, \quad V^l_{ps} = Y^l_{ps} W_{ps,V}^{l},
\\
A^l_{ps} = \text{softmax} \left( \frac{Q^l_{ps} {K^l_{ps}}^{\top}}{\sqrt{d_k}} \right),
\\
O_{ps,h}^{l} = A^l_{ps} V^l_{ps},
\\
O^{l}_{ps,mhsa} = \text{Concat}(O_{1}^{l}, O_{2}^{l}, \dots, O_{H}^{l}) W^l_O.
\end{array}
\label{eq:mhsa}
\end{equation}
\( Y^l_{ps} \) represents the input feature; \( W_{ps,Q}^{l}\), \(W_{ps,K}^{l}\), and \(W_{ps,V}^{l} \) are learnable weight matrices that project the input \( Y^l_{ps} \) into the query (\( Q^l_{ps} \)), key (\( K^l_{ps} \)), and value (\( V^l_{ps} \)) matrices, respectively. The attention matrix \( A^l_{ps} \) is calculated using the scaled dot-product of the queries and keys, scaled by the square root of the key dimension \( d_k \). The attention output for the \( h \)-th attention head, \( O_{ps,h}^{l} \), is calculated as the product of the attention weights \( A^l_{ps} \) and the value matrix \( V^l_{ps} \). The outputs \( O_{1}^{l}, O_{2}^{l}, \dots, O_{H}^{l} \) correspond to the results of \( H \) attention heads, computed in parallel. Finally, \( W_O^l \) is the final projection matrix that combines the outputs of all attention heads. Following the operation described in Eq. (\ref{eq:mhsa}), by inputting \( Y^l_{fb} \) and \( Y^l_{ws} \) with their corresponding attention weights, we can obtain \( O^{l}_{fb,mhsa} \) and \( O^{l}_{ws,mhsa} \), respectively.

In the subsequent steps of FiDo, the concatenation of 
$O^{l}_{ps,mhsa}$ and $O^{l}_{fb,mhsa}$ is processed through feature-based concatenation, followed by projecting the concatenated features into the CNN module for higher-level projection steps. The detailed process is as follows.

\begin{equation}
\begin{array}{c}
O_{concat}^{l} = \text{Concat}(O^{l}_{ps,mhsa}, O^{l}_{fb,mhsa}) \in \mathbb{R}^{T \times (d_{ps} + d_{fb})},\\
O_{CNN}^l = \text{CNN}(O_{concat}^{l}).

\end{array}
\end{equation}
In the final stage when FiDo obtains \(O^l_{\text{FiDo}} \in \mathbb{R}^{T \times (d_{\text{cnn}} + d_{\text{adapt.}})}\), $O^l_{CNN}$ is concatenated with the adapter-processed $O^{l}_{ws,mhsa}$. The detailed process is as follows:

\begin{equation}
\begin{array}{c}
%O^l_{\text{FiDo}} \in \mathbb{R}^{T \times (d_{\text{cnn}} + d_{\text{adapt.}})}, 
O^l_{\text{FiDo}} = \text{Concat}(O^l_{\text{CNN}}, \text{Adapter}(O_{\text{ws,mhsa}}^{\text{l}})).
\end{array}
\end{equation}
$O^r_{\text{FiDo}}$ for the right channel is obtained through the same steps. $O^l_{\text{FiDo}}$ and $O^r_{\text{FiDo}}$ are then used as input to the assessment modules, where we adopt the original implementation of MBI-Net+ \cite{zezario24_interspeech}. Specifically, BLSTM+AT is applied to obtain frame-level intelligibility and HASPI scores and the embedding representation input to the class module (HA class). The frame-level intelligibility and HASPI scores are processed by their corresponding task-specific modules to predict the final intelligibility and HASPI scores, respectively. The loss function for training MBI-Net+ with FiDo is defined as follows:

\begin{equation}
\label{eq:overall_loss}
   \small
    \begin{array}{c}
    O = \gamma_{1}\L_{Int} + \gamma_{2}\L_{HASPI} + \gamma_{3}\L_{CE},
    \end{array} 
\end{equation}
where $\L_{CE}$ represents the cross-entropy loss between the estimated enhancement type of the class module and the reference enhancement type (the front-end processor in the HA system). The weights between the losses are represented by $\gamma_1$, $\gamma_2$, and $\gamma_3$. $\L_{Int}$ and $\L_{HASPI}$ represent the MSE losses between the corresponding ground-truth scores and the predicted scores.

\section{Experiments}

\subsection{Experimental Setup}
The Clarity Prediction Challenge (CPC) 2023 dataset \cite{barker2024} includes recordings from six talkers, ten enhancement methods (corresponding to ten hearing aid (HA) systems) from the 2022 Clarity Enhancement Challenge \cite{10094918}, and intelligibility ratings provided by 25 listeners. The dataset is categorized into three tracks: Track 1 contains 2,779 utterances, Track 2 has 2,796 utterances, and Track 3 consists of 2,772 utterances. For each track, 90\% of the utterances were allocated for training, while the remaining 10\% were used for development. The test sets for the three tracks comprise 305, 294, and 298 utterances, respectively, featuring unseen listeners and HA systems. The models were trained and evaluated separately on all three tracks, and the overall performance across them is referred to as "All." The evaluation relied on three key metrics: root mean square error (RMSE), linear correlation coefficient (LCC), and Spearman's rank correlation coefficient (SRCC) \cite{srcc}. A lower RMSE signifies better alignment with ground-truth scores, while higher LCC and SRCC values indicate stronger correlations between predictions and actual scores. All models presented in this study were trained and assessed using the CPC 2023 dataset.

\subsection{Comparison of different feature concatenation strategies}

In the first experiment, we aim to compare the performance of MBI-Net+ using different feature concatenation strategies. This experiment intends to confirm the most appropriate concatenation method when employing cross-domain features.

Following the original implementation of MBI-Net+ \cite{zezario24_interspeech}, we adopt a structure consisting of a cross-domain feature extraction module, a four-layer convolutional block (with 16, 32, 64, and 128 channels, respectively), a one-layer Bidirectional Long Short-Term Memory (BLSTM) (with 128 nodes), and a fully connected layer (with 128 neurons). The model also incorporates an attention mechanism and a single neuron for frame-level scoring. Finally, two task-specific modules are leveraged for estimating subjective intelligibility and HASPI, along with a classification module trained to classify ten enhancement system types. The model is trained using the Adam \cite{KingmaB14} optimizer with a learning rate of 1e-4. We set the tunable weights to $\gamma_{1}$ = 1,  $\gamma_{2}$ = 0.4, and  $\gamma_{3}$ = 0.2, assigning higher weights to more important metrics and ensuring that the losses remain on a similar scale. Specifically, we deployed three versions of MBI-Net+: (1) MBI-Net+ with the Whisper Medium model using temporal concatenation, (2) MBI-Net+ with the Whisper Large model using temporal concatenation, and (3) MBI-Net+ with the Whisper Large model using feature-based concatenation. For temporal concatenation, the frames of PS, FB, and WS are combined, resulting in a frame size equal to the sum of their frame numbers.

\begin{figure}[t]
\centering
\includegraphics[width=8.3cm]{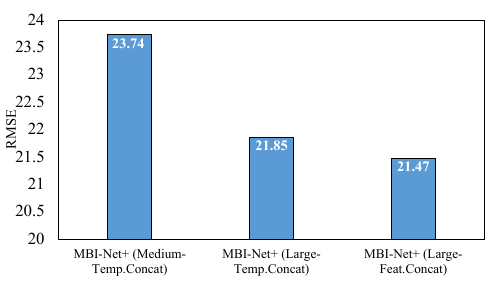} 
\caption{Performance comparison of different feature concatenation strategies.} 
\label{fig:concat}
\end{figure}

In this experiment, we specifically focus on Track 3, which has a smaller training set, to evaluate the feature concatenation strategy. As shown in Fig. \ref{fig:concat}, we first notice that by simply replacing Whisper Medium (the original implementation of MBI-Net+ \cite{zezario24_interspeech}) with Whisper Large, we can achieve a relative RMSE reduction of 7.96\% (from 23.74 to 21.85). This confirms that feature extraction plays an important role in improving prediction performance. Furthermore, replacing temporal concatenation with feature-based concatenation further improves the RMSE score. Therefore, this confirms that feature-based concatenation provides an effective strategy while also avoiding the lengthy number of frames present in temporal concatenation.

\graphicspath{ {./images/} }

\graphicspath{ {./images/} }
\begin{figure}[t]
\centering
\includegraphics[width=8.2 cm]{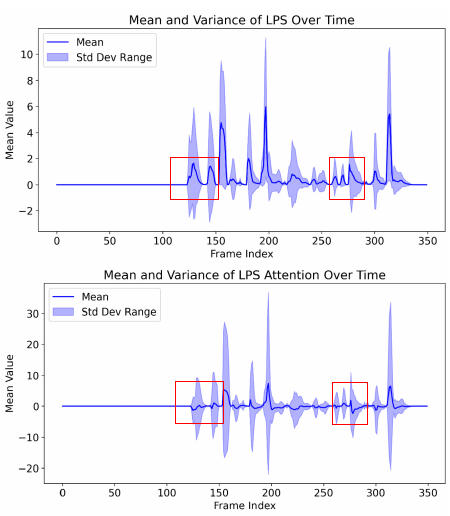} 
\caption{Mean and standard deviation analysis between PS features before and after performing the attention mechanism.} 
\label{fig:corr}
\end{figure}

\subsection{Comparing MBI-Net+ with FiDo and MBI-Net+}

In the second experiment, we investigate the impact of FiDo on the performance of MBI-Net+ across different tracks. FiDo employs eight attention heads for each input feature: PS, FB, and WS. After applying attention, PS and FB features are concatenated. This concatenated output is then processed by a four-layer convolutional block with 16, 32, 64, and 128 channels, respectively. For the MHSA output of WS features, an adapter layer is introduced to perform task-specific adaptation and reduce the dimensionality to align with the output of the convolutional block. Finally, the outputs of the convolutional block and the adapter layer are concatenated and fed into the BLSTM attention layer and the corresponding assessment module, following the original MBI-Net+ configuration \cite{zezario24_interspeech}.

As shown in Table 1, for Track 1, FiDo achieves an LCC of 0.754 and an SRCC of 0.728, reducing the RMSE from 28.370 to 26.710. In Track 2, the improvements are more pronounced, with LCC increasing from 0.754 to 0.791 and SRCC improving from 0.737 to 0.787, while RMSE decreases from 25.920 to 23.850. In Track 3, applying FiDo boosts the LCC from 0.813 to 0.850, improves the SRCC from 0.814 to 0.861, and reduces the RMSE from 23.740 to 21.420. Overall, across all tracks, FiDo improves LCC from 0.764 to 0.796, SRCC from 0.767 to 0.793, and reduces RMSE from 26.100 to 24.110. These results confirm that FiDo’s early attention mechanism and feature-based concatenation strategy help improve the model’s prediction performance.

In the next experiment, we visualize the effect of attention by comparing the mean and standard deviation of PS features before and after applying the attention mechanism. As shown in Fig. \ref{fig:corr}, the attention mechanism changes the shape of the feature, which further confirms that the attention mechanism indeed emphasizes specific regions of the audio representation.

\subsection{Comparison with Other Prediction Models for HA}
In the last experiment, we further compare MBI-Net+ with FiDo against other state-of-the-art systems on the Clarity Prediction Challenge 2023 dataset. The results are shown in Table 2. The first observation is that models employing large-scale pre-trained models (E011 \cite{cuervo2024speech}, E002 \cite{mogridge2024nonintrusive}, E025 \cite{barker2024}, and MBI-Net+ \cite{zezario2022mbi}) generally outperform models that rely on conventional signal processing techniques (E003, E024, E015, and E020). Interestingly, when compared to intrusive-based systems, three models, including MBI-Net+ with FiDo, which employs Whisper as the feature extraction module, achieve lower RMSE values than the intrusive models. Furthermore, MBI-Net+ with FiDo achieves a relative RMSE reduction of 3.98\% (from 25.1 to 24.1) compared to the best-performing model E011 \cite{cuervo2024speech} in the 2023 Clarity Prediction Challenge, which adopts a more complex and deeper model architecture. This further confirms that by extracting acoustic features more effectively and focusing on important regions at an early stage, the model can achieve better prediction performance.

\begin{table}[t]
\caption{LCC, SRCC, and RMSE results of MBI-Net+ and MBI-Net+ with FiDo on the complete test set.}
\footnotesize
\begin{center}
 \begin{tabular}{c||c||c||c} 
 \hline
 \hline
 \textbf{Model} &\textbf{LCC} & \textbf{SRCC} & \textbf{RMSE}  \\ [0.5ex] \cline{2-4}
 \hline\hline
  \multicolumn{4}{c} {Track 1}
\\ \hline
MBI-Net+ \cite{zezario24_interspeech} &0.721&0.714&28.370\\\hline
MBI-Net+ with FiDo&\textbf{0.754}&\textbf{0.728}&\textbf{26.710}\\\hline
 \hline
  \multicolumn{4}{c} {Track 2} \\
 \hline
MBI-Net+ \cite{zezario24_interspeech} &0.754&0.737&25.920\\\hline
MBI-Net+ with FiDo &\textbf{0.791}&\textbf{0.787}&\textbf{23.850}\\\hline
 \hline
  \multicolumn{4}{c} {Track 3} \\
 \hline
MBI-Net+ \cite{zezario24_interspeech} &0.813&0.814&23.740\\\hline
MBI-Net+ with FiDo&\textbf{0.850}&\textbf{0.861}&\textbf{21.420}\\\hline
 \hline
  \multicolumn{4}{c} {Track All} \\
 \hline
MBI-Net+ \cite{zezario24_interspeech}  &0.764&0.767&26.100\\\hline
MBI-Net+ with FiDo &\textbf{0.796}&\textbf{0.793}&\textbf{24.110}\\\hline
 \hline
\end{tabular}
\end{center}
\end{table}

\begin{table}[t]
\caption{RMSE and LCC results of different systems (Intrusive and Non-Intrusive) from the First and Second Clarity Challenges on the test set.}
\footnotesize
\begin{center}
 \begin{tabular}{c||c||c||c} 
 \hline
 \hline
 \textbf{System} & \textbf{Non-Intrusive} & \textbf{RMSE} & \textbf{LCC}  \\ [0.5ex] 
 \hline
\hline
MBI-Net+ with FiDo&\checkmark&\textbf{24.1}&\textbf{0.80} \\ \hline 
E011 \cite{cuervo2024speech}&\checkmark&25.1&0.78 \\ \hline 
E002 \cite{mogridge2024nonintrusive}&\checkmark&25.3&0.77 \\ \hline 
E009 \cite{Huckvale} &\ding{53}&25.4&0.78 \\ \hline 
E022 \cite{tu_barker} &\ding{53}&25.7&0.77 \\ \hline 
MBI-Net+ \cite{zezario24_interspeech}&\checkmark&26.1&0.76 \\ \hline
%MBI-Net+&Yes&26.8&0.75 \\ \hline
%MBI-Net+&Yes&26.8&0.754 \\ \hline
E025 \cite{tu_barker} &\checkmark&27.9&0.72 \\ \hline 
Baseline \cite{barker2024}&\ding{53}&28.7&0.70 \\ \hline 
E003 \cite{mawalim_isca} &\checkmark&31.1&0.64 \\ \hline 
E024 \cite{mawalim_isca} &\checkmark&31.7&0.62 \\ \hline 
E015 \cite{yamamoto_} &\checkmark&35.0&0.60 \\ \hline 
E020 \cite{mamum} &\checkmark&39.8&0.33 \\ \hline 
Prior \cite{barker22_interspeech}&\ding{53}&40 & - \\ \hline 
 \hline

\end{tabular}
\end{center}
\end{table}

\section{Conclusion}
In this paper, we have proposed Feature Importance across Domains (FiDo), which integrates multi-head self-attention (MHSA) and feature-based concatenation to capture the importance of different acoustic features across multiple domains. Experimental results confirm that incorporating FiDo into MBI-Net+ improves overall prediction performance, achieving a substantial RMSE reduction of 7.62\% over the original MBI-Net+ model. Furthermore, MBI-Net+ with FiDo achieves a 3.98\% RMSE reduction over the top system in the 2023 Clarity Prediction Challenge. These results validate the effectiveness of FiDo in enhancing speech intelligibility assessment in hearing aids and highlight the importance of capturing more effective acoustic features. In future work, we will focus on further refining FiDo’s feature extraction process by exploring advanced attention mechanisms and feature fusion strategies.

\bibliographystyle{IEEEtran}
\bibliography{mybib}

\begin{thebibliography}{10}
\providecommand{\url}[1]{#1}
\csname url@samestyle\endcsname
\providecommand{\newblock}{\relax}
\providecommand{\bibinfo}[2]{#2}
\providecommand{\BIBentrySTDinterwordspacing}{\spaceskip=0pt\relax}
\providecommand{\BIBentryALTinterwordstretchfactor}{4}
\providecommand{\BIBentryALTinterwordspacing}{\spaceskip=\fontdimen2\font plus
\BIBentryALTinterwordstretchfactor\fontdimen3\font minus \fontdimen4\font\relax}
\providecommand{\BIBforeignlanguage}[2]{{%
\expandafter\ifx\csname l@#1\endcsname\relax
\typeout{** WARNING: IEEEtran.bst: No hyphenation pattern has been}%
\typeout{** loaded for the language `#1'. Using the pattern for}%
\typeout{** the default language instead.}%
\else
\language=\csname l@#1\endcsname
\fi
#2}}
\providecommand{\BIBdecl}{\relax}
\BIBdecl

\bibitem{loizou2007speech}
P.~C. Loizou, \emph{Speech Enhancement: Theory and Practice}.\hskip 1em plus 0.5em minus 0.4em\relax CRC press, 2007.

\bibitem{10248166}
G.~Close, T.~Hain, and S.~Goetze, ``{The Effect of Spoken Language on Speech Enhancement Using Self-Supervised Speech Representation Loss Functions},'' in \emph{Proc. WASPAA}, 2023, pp. 1--5.

\bibitem{barker22_interspeech}
J.~Barker, M.~Akeroyd, J.~Trevor, J.~Culling, J.~Firth, S.~Graetzer, H.~Griffiths, L.~Harris, G.~Naylor, Z.~Podwinska, E.~Porter, and R.~Munoz, ``{The 1st Clarity Prediction Challenge: A Machine Learning Challenge for Hearing Aid Intelligibility Prediction},'' in \emph{Proc. INTERSPEECH}, 2022, pp. 3508--3512.

\bibitem{10362991}
B.~Schuh, W.~Wardah, B.~Naderi, T.~Michal, and S.~Moeller, ``{Hearing Impairment in Crowdsourced Speech Quality Assessments: Its Effect and Screening with Digit Triplet Hearing Test},'' in \emph{Speech Communication; 15th ITG Conference}, 2023, pp. 21--25.

\bibitem{ref_36}
{ANSI Std. S3.5 1997}, ``{Methods for Calculation of The Speech Intelligibility Index},'' in \emph{Acoustical Society of America}, 1997.

\bibitem{ref_37}
T.~Houtgast and H.~.~M. Steeneken, ``{Evaluation of Speech Transmission Channels by Using Artificial Signals},'' \emph{Acustica}, vol.~25, no.~6, pp. 355--367, 1971.

\bibitem{ref_38}
H.~J.~M. Steeneken and T.~Houtgast, ``{A Physical Method for Measuring Speech-Transmission Quality},'' \emph{Journal of the Acoustical Society of America}, vol.~67, no.~1, pp. 318--326, 1980.

\bibitem{ref_39}
C.~H. Taal, R.~C. Hendriks, R.~Heusdens, and J.~Jensen, ``{An Algorithm for Intelligibility Prediction of Time-frequency Weighted Noisy Speech},'' \emph{IEEE/ACM Transactions on Audio, Speech and Language Processing}, vol.~19, no.~7, pp. 2125--2136, 2011.

\bibitem{ANDERSEN20181}
A.~H. Andersenan, J.~M. Haan, Z.-H. Tan, and J.Jensen, ``{Refinement and Validation of The Binaural Short Time Objective Intelligibility Measure for Spatially Diverse Conditions},'' \emph{Speech Communication}, vol. 102, pp. 1--13, 2018.

\bibitem{katehaspi}
J.~M. Kates and K.~H. Arehart, ``{The Hearing-Aid Speech Perception Index ({HASPI}) Version 2},'' \emph{Speech Communication}, vol. 131, pp. 35--46, 2021.

\bibitem{ssl-mos}
E.~Cooper, W.-H. Huang, T.~Toda, and J.~Yamagishi, ``{Generalization Ability of MOS Prediction Networks},'' in \emph{Proc. ICASSP}, 2022, pp. 8442--8446.

\bibitem{mosa-net}
R.~Zezario, S.-W. Fu, F.~Chen, C.-S. Fuh, H.-M. Wan, and Y.~Tsao, ``{Deep Learning-Based Non-Intrusive Multi-Objective Speech Assessment Model With Cross-Domain Features},'' \emph{IEEE/ACM Transactions on Audio, Speech, and Language Processing}, vol.~31, pp. 54--70, 2023.

\bibitem{zezario2024studyincorporatingwhisperrobust}
R.~E. Zezario, Y.-W. Chen, S.-W. Fu, Y.~Tsao, H.-M. Wang, and C.-S. Fuh, ``A study on incorporating {Whisper} for robust speech assessment,'' in \emph{Proc. ICME}, 2024.

\bibitem{yang22o_interspeech}
Z.~Yang, W.~Zhou, C.~Chu, S.~Li, R.~Dabre, R.~Rubino, and Y.~Zhao, ``{Fusion of Self-Supervised Learned Models for MOS Prediction},'' in \emph{Proc. INTERSPEECH}, 2022, pp. 5443--5447.

\bibitem{cuervo2024speech}
S.~Cuervo and R.~Marxer, ``{Speech Foundation Models on Intelligibility Prediction for Hearing-Impaired Listeners},'' in \emph{Proc. ICASSP}, 2024, pp. 1421--1425.

\bibitem{mogridge2024nonintrusive}
R.~Mogridge, G.~Close, R.~Sutherland, T.~Hain, J.~Barker, S.~Goetze, and A.~Ragni, ``{Non-Intrusive Speech Intelligibility Prediction for Hearing-Impaired Users Using Intermediate ASR Features and Human Memory Models},'' in \emph{Proc. ICASSP}, 2024, pp. 306--310.

\bibitem{10094918}
M.~A. Akeroyd, W.~Bailey, J.~Barker, T.~J. Cox, J.~F. Culling, S.~Graetzer, G.~Naylor, Z.~Podwińska, and Z.~Tu, ``{The 2nd Clarity Enhancement Challenge for Hearing Aid Speech Intelligibility Enhancement: Overview and Outcomes},'' in \emph{Proc. ICASSP}, 2023, pp. 1--5.

\bibitem{barker2022is}
J.~Barker, M.~Akeroyd, T.~J. Cox, J.~F. Culling, J.~Firth, S.~Graetzer, H.~Griffiths, L.~Harris, G.~Naylor, Z.~Podwinska, E.~Porter, and R.~V. Munoz, ``The 1st clarity prediction challenge: A machine learning challenge for hearing aid intelligibility prediction,'' in \emph{Proc. INTERSPEECH}, 2022.

\bibitem{zezario24_interspeech}
R.~E. Zezario, F.~Chen, C.-S. Fuh, H.-M. Wang, and Y.~Tsao, ``Non-intrusive speech intelligibility prediction for hearing aids using whisper and metadata,'' in \emph{Proc. INTERSPEECH}, 2024, pp. 3844--3848.

\bibitem{chiang2023multiobjective}
H.-T. Chiang, S.-W. Fu, H.-M. Wang, Y.~Tsao, and J.~H.~L. Hansen, ``{Multi-objective Non-intrusive Hearing-aid Speech Assessment Model},'' \emph{J. Acoust. Soc. Am.}, vol. 195, p. 3574–3587, 2024.

\bibitem{tu22_interspeech}
Z.~Tu, N.~Ma, and J.~Barker, ``{Exploiting Hidden Representations from a DNN-based Speech Recogniser for Speech Intelligibility Prediction in Hearing-Impaired Listeners},'' in \emph{Proc. INTERSPEECH}, 2022, pp. 3488--3492.

\bibitem{MAWALIM2023109663}
C.~O. Mawalim, B.~A. Titalim, S.~Okada, and M.~Unoki, ``{Non-Intrusive Speech Intelligibility Prediction Using an Auditory Periphery Model with Hearing Loss},'' \emph{Applied Acoustics}, vol. 214, p. 109663, 2023.

\bibitem{zezario2022mbi}
R.~E. Zezario, F.~Chen, C.~S. Fuh, H.-M. Wang, and Y.~Tsao, ``Mbi-net: a non-intrusive multi-branched speech intelligibility prediction model for hearing aids,'' in \emph{Proc. INTERSPEECH}, 2022, pp. 3944--3948.

\bibitem{9709938}
K.~H. Lee, C.~Park, J.~Oh, and N.~Kwak, ``{ LFI-CAM: Learning Feature Importance for Better Visual Explanation },'' in \emph{Proc. IEEE/CVF ICCV}, 2021, pp. 1335--1343.

\bibitem{NIPS2017_3f5ee243}
A.~Vaswani, N.~Shazeer, N.~Parmar, J.~Uszkoreit, L.~Jones, A.~N. Gomez, L.~u. Kaiser, and I.~Polosukhin, ``Attention is all you need,'' in \emph{Advances in Neural Information Processing Systems}, vol.~30, 2017.

\bibitem{sincnet}
M.~Ravanelli and Y.~Bengio, ``Speaker recognition from raw waveform with {SincNet},'' in \emph{Proc. SLT}, no. 1021-1028, 2018.

\bibitem{barker2024}
J.~P. Barker, M.~A. Akeroyd, W.~Bailey, T.J.Cox, J.~F. Culling, J.~Firth, S.~Graetzer, and G.~Naylor, ``{The 2nd Clarity Prediction Challenge: A Machine Learning Challenge for Hearing Aid Intelligibility Prediction},'' in \emph{Proc. ICASSP}, 2024, pp. 11\,551--11\,555.

\bibitem{srcc}
C.~Spearman, ``{The Proof and Measurement of Association between Two Things},'' \emph{The American Journal of Psychology}, vol.~15, no.~1, pp. 72--101, 1904.

\bibitem{KingmaB14}
D.~Kingma and J.Ba, ``{Adam: A Method for Stochastic Optimization},'' in \emph{Proc. ICLR}, 2015, pp. 1--13.

\bibitem{Huckvale}
M.~Huckvale and G.~Hilkhuysen, ``Combining acoustic, phonetic, linguistic and audiometric data in an intrusive intelligibility metric for hearing-impaired listeners,'' in \emph{Proc. ISCA Clarity 2023}, 2023.

\bibitem{tu_barker}
Z.~Tu, N.~Ma, and J.~Barker, ``Intelligibility prediction with a pretrained noise-robust automatic speech recognition model,'' in \emph{Proc. ISCA Clarity 2023}, 2023.

\bibitem{mawalim_isca}
C.~O. Mawalim, X.~Zhou, S.~Okada, and M.~Unoki, ``A nonintrusive speech intelligibility prediction using binaural cues and time-series model with one-hot listener embedding,'' in \emph{Proc. ISCA Clarity 2023}, 2023.

\bibitem{yamamoto_}
K.~Yamamoto, ``A non-intrusive binaural speech intelligibility prediction for clarity-2023,'' in \emph{Proc. ISCA Clarity 2023}, 2023.

\bibitem{mamum}
N.~Mamun, S.~Ahmed, and J.~H.~L. Hansen, ``Prediction of behavioral speech intelligibility using a computational model of the auditory system,'' in \emph{Proc. ISCA Clarity 2023}, 2023.

\end{thebibliography}

% \begin{thebibliography}{9}
% \bibitem[1]{Davis80-COP}
%   S.\ B.\ Davis and P.\ Mermelstein,
%   ``Comparison of parametric representation for monosyllabic word recognition in continuously spoken sentences,''
%   \textit{IEEE Transactions on Acoustics, Speech and Signal Processing}, vol.~28, no.~4, pp.~357--366, 1980.
% \bibitem[2]{Rabiner89-ATO}
%   L.\ R.\ Rabiner,
%   ``A tutorial on hidden Markov models and selected applications in speech recognition,''
%   \textit{Proceedings of the IEEE}, vol.~77, no.~2, pp.~257-286, 1989.
% \bibitem[3]{Hastie09-TEO}
%   T.\ Hastie, R.\ Tibshirani, and J.\ Friedman,
%   \textit{The Elements of Statistical Learning -- Data Mining, Inference, and Prediction}.
%   New York: Springer, 2009.
% \bibitem[4]{YourName17-XXX}
%   F.\ Lastname1, F.\ Lastname2, and F.\ Lastname3,
%   ``Title of your INTERSPEECH 2021 publication,''
%   in \textit{Interspeech 2021 -- 20\textsuperscript{th} Annual Conference of the International Speech Communication Association, September 15-19, Graz, Austria, Proceedings, Proceedings}, 2020, pp.~100--104.
% \end{thebibliography}

\end{document}